\documentstyle[11pt,xray2000,twoside,epsf]{article}
\reversemarginpar

\begin{document}
\title{Detecting X-ray Synchrotron Emission in Supernova Remnants: Implications for Abundances and Cosmic Rays}
 \author{Kristy K. Dyer, Stephen P. Reynolds, Kazik J. Borkowski}
\affil{North Carolina State University, Physics Department Box 8202, Raleigh NC 27695}
\author{Robert Petre}
\affil{NASA's GSFC, LHEA Code 666, Greenbelt MD 20771}

\begin{abstract}
The 10$^{51}$ ergs released in a supernova have far reaching consequences in
the galaxy, determining elemental abundances, accelerating cosmic rays,
and affecting the makeup of the interstellar medium. Recently the spectra
of several supernova remnants have been found to be dominated by nonthermal
emission. Separating the thermal and nonthermal components is
important not only for the understanding of cosmic-ray acceleration and
shock microphysics properties but for accurate assessment of the
temperatures and line strengths. New models designed
to model spatially resolved synchrotron X-rays from type Ia supernovae can contribute to the understanding of both the thermal physics (dynamics, abundances) and nonthermal physics (shock acceleration, magnetic-field amplification) of
supernova remnants. I will describe model fits to SN 1006,
emphasizing the physical constraints that can be placed on SNRs,
abundances, and the cosmic-ray acceleration process.
\end{abstract}

\section{Introduction}

\begin{figure} 
\plotfiddle{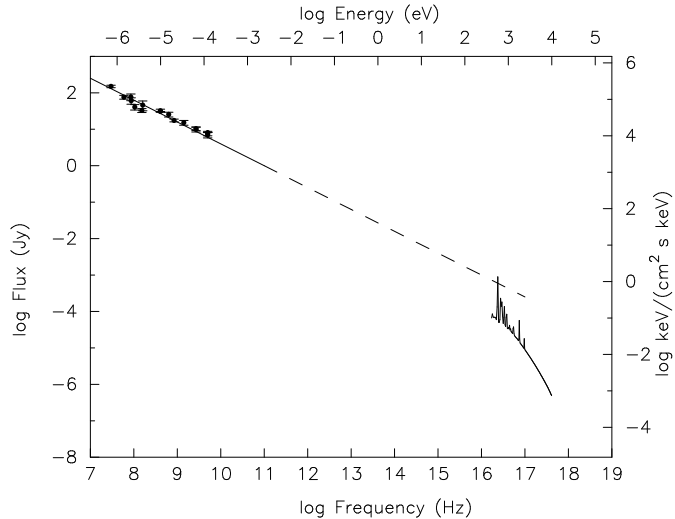}{2.5in}{0}{50}{50}{-150}{-170}
\caption{The radio flux measurement with a power law spectrum extrapolated to X-ray energies. The X-rays are the unfolded ASCA spectrum, shown for comparison.
 \label{fluxes}} 
\end{figure}

A growing number of SNRs (listed in Table 1) are thought to show evidence for synchrotron X-rays, for a variety of reasons. Most are suspected of nonthermal X-rays because they show hard energy excesses that can be fit with powerlaws. A few  (SN~1006, G347.3-0.5) have nearly featureless X-ray spectra which are clearly dominated by nonthermal emission. Some remnants, while not dominated by nonthermal emission in the ASCA band, show other evidence pointing to its presence. OSSE observations of Cassiopeia A at 400-1250 keV (The 1996) are well described by a broken power law, steepening to higher energies (Allen 1997).
RCW 86 is an especially interesting case. It shows anomalously low abundances when fit with thermal models (Vink 1997), but the addition of a synchrotron model bring the abundances to within expected ranges (Borkowski 2001).


The most straightforward explanation for nonthermal X-ray emission (non-plerionic in these SNRs) is synchrotron emission from the same mechanism that is responsible for the radio emission.
It is straightforward to determine the synchrotron spectrum from the electron energy population 
However, the extrapolation from the radio spectrum measured by single dish instruments, over-predicts the X-ray flux (see Figure 1) -- there must be a mechanism that causes the electron population to roll off at the highest energies. Reynolds \& Chevalier (1981) attributed this rolloff to synchrotron losses.

The sharpest physically plausible spectral rolloff would be the
emission resulting from an exponential cutoff of the electron
power-law energy distribution, radiating in a uniform magnetic field.  
This was embodied in the XSPEC model {\it SRCUT} and used by Reynolds \& Keohane (1999) to set limits on the energies to which SNRs could accelerate electrons (with implications for cosmic ray ions). A more sophisticated treatment requires addressing the mechanism limiting electrons at the highest energies. The cutoff could be due to the age of the SNR, radiative losses from synchrotron or inverse-Compton emission, or electron escape(Reynolds 1998).


\begin{table}
\caption{SNRs which have reported nonthermal X-rays\label{table1}}
\begin{tabular}{lcc}
\tableline
SNR & Instrument & Nonthermal reference \\
\tableline
Cassiopeia A	& OSSE, RXTE, BeppoSAX & The 1996, Allen 1997, Favata 1997\\
RCW 86		& ASCA				& Borkowski 2001, Bamba 2001\\
SN 1006		& ASCA			& Koyama 1995, Dyer 2001\\
Kepler 		& ASCA, XTE			& Decourchelle 1999\\
G347.3-0.5	& ASCA 				& Koyama 1997, Slane 1999\\
G327.1-1.1	& ASCA, ROSAT			& Sun 1999\\ 
\tableline
\tableline
\end{tabular}
\end{table}

%
%

\section{The Escape-limited Synchrotron Model}
\begin{figure} 
\plotfiddle{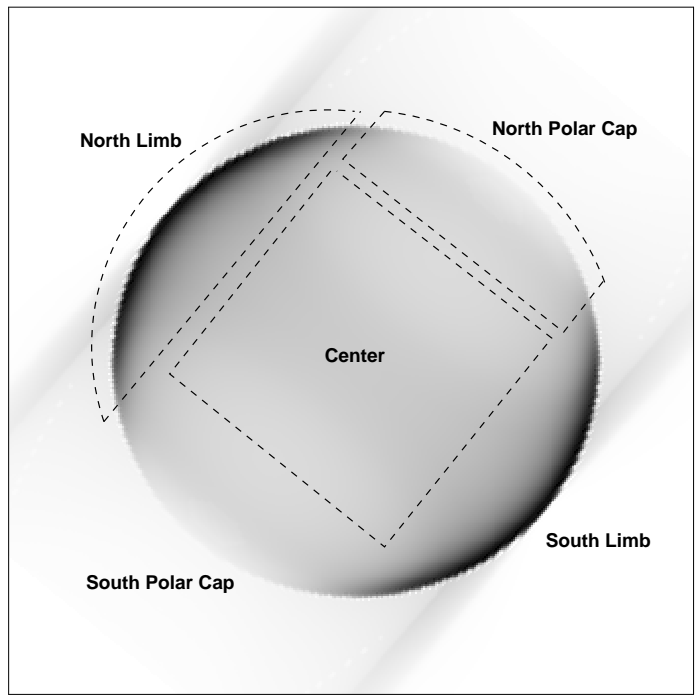}{275pt}{0}{50}{50}{-150}{0}
\caption{Spatially resolved subsets of the {\it SRESC} model developed for SN~1006.
 \label{regions}} 
\end{figure}
The escape-limited model is a single parameter model that is particularly well suited to X-ray analysis programs such as XSPEC, and has been released as {\it SRESC} in XSPEC 11 (it should also be available for Chandra data analysis in CIAO 2.1). This model uses the Sedov solution to specify the electron distribution and assumes a uniform upstream magnetic field, with a shock compression ratio of 4. The model takes the radio spectral index and 1 GHz flux as inputs (available from Green 2000 for galactic SNRs and now for some Magellanic Cloud SNRs in Williams 2000) and fits one parameter, the rolloff frequency --  proportional to the product of the maximum MHD wavelength scattering the electrons and the magnetic field strength. We have applied {\it SRESC} to archival observations of the integrated ASCA spectrum of SN 1006 as a test case. As well as describing the physics behind the emission the SRESC model is much better constrained for fitting purposes than a powerlaw model. The synchrotron model allows more accurate abundance determinations in the thermal emission -- these abundances are close to predictions for Type Ia SN (Iwamoto et al. 1999) and include half a solar mass of iron. Since TeV detections can be used to determine the relativistic electron density in SN 1006, fitting the rolloff frequency with the SRESC model uniquely determines the magnetic field. This suggests that SN~1006 is in fact far from equipartition, violating a common assumption for astrophysical plasmas. The model also makes spatial predictions and is capable of reproducing the morphology of SN 1006 at energies at which it is dominated by nonthermal emission, shown in Figure 2.

However, a much more stringent test of the model is to apply it to spatially resolved spectra from SN 1006. As shown in Figure 2 we have developed subsets of the escape-limited synchrotron model for different regions of the supernova remnant, describing the polar caps, the limbs and the center. These models predict slightly different curvature for the nonthermal emission in each region for a particular rolloff energy. The models should allow us to make more accurate determinations of the abundances as well as mapping the thermal and nonthermal distribution across the remnant.

Preliminary results have been very promising. Contrary to expectation, all regions of SN 1006 have a significant amount of nonthermal emission, including the polar caps and center which have been fit in the past with thermal-only models. The limbs are dominated by nonthermal emission but require a thermal
component while fits to the thermally dominated center will tolerate a significant amount of nonthermal emission. In addition we find that the north limb and south limb will not support the identical thermal and nonthermal models -- we find the spectra to be detectably different.


\section{Conclusions} 


We have found, both in integrated and spatially resolved models, that the measured abundances and temperature change with the addition of accurate synchrotron models. In the future, accurate accounting for a possible nonthermal component will be a necessary precondition to using nucleosynthesis model predictions to interpret abundances in any SNR. {\it SRCUT} and {\it SRESC}, new models in XSPEC 11, accurately describe this emission and are a substantial improvement over powerlaw models both in terms of physical description and fitting.


All regions of SN 1006, even the polar caps which show obvious thermal lines, seem to have significant nonthermal emission. We also find that the spectra of the north and the south limb are not identical. While most of the emission from the center does seem to be thermal, the fit can tolerate a significant amount of nonthermal emission.

Technology is surpassing theory -- Chandra arcsecond-resolution observations of SNRs including SN~1006 will soon be available, requiring models capable of describing extremely small scale variations. These synchrotron models will be useful both for analyzing higher resolution observations of SN~1006, and for interpreting other SNRs with nonthermal X-ray emission.

\begin{figure} 
\plotfiddle{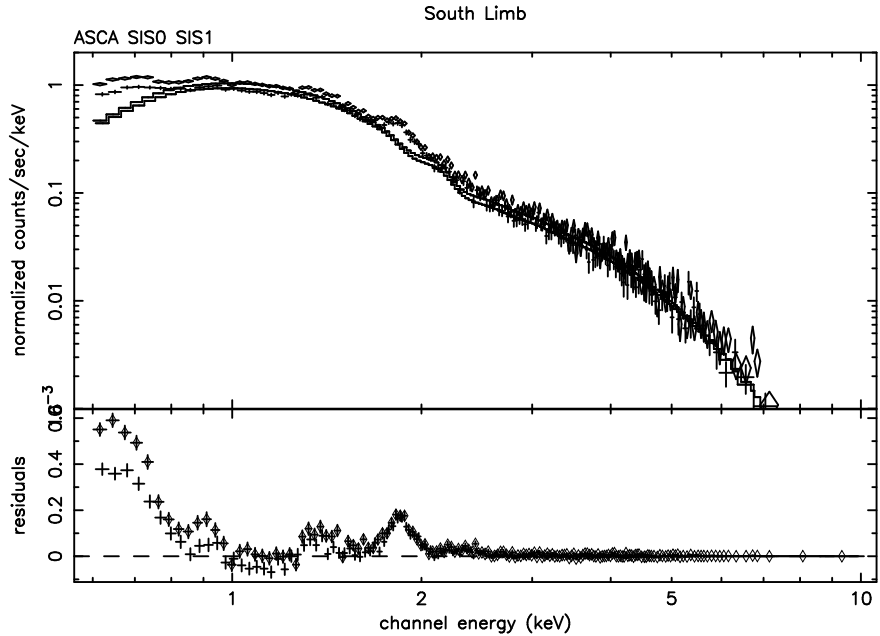}{237pt}{-90}{50}{50}{-190}{275}
\caption{A limb version of the SRESC model fit to ASCA data from the south limb of SN~1006. Note the line-like residuals.
\label{needthermal}} 
\end{figure}

\begin{figure} 
\plotfiddle{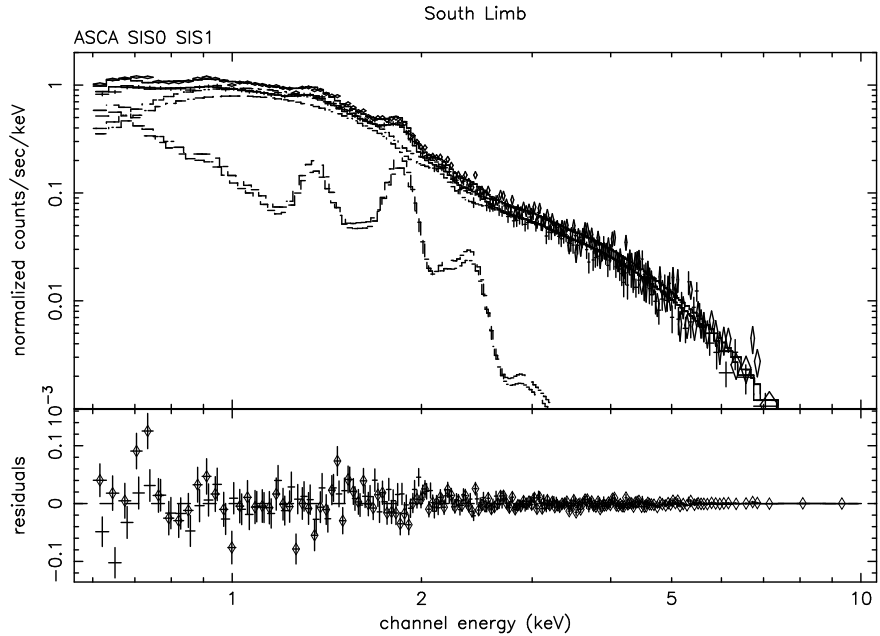}{237pt}{-90}{50}{50}{-190}{275}
\caption{The limb version of the SRESC model fit to data from the south limb of SN~1006 with a plane shock (VPSHOCK, Borkowski 2001).
\label{thermallimb}} 
\end{figure}

\acknowledgments

This work is supported by NASA's Graduate Student Research Program through GSFC. {\it http://education.nasa.gov/gsrp/ } 


\end{document}